# Spontaneous decays of magneto-elastic excitations in noncollinear antiferromagnet (Y,Lu)MnO$_3$

Joosung Oh[1,2], Manh Duc Le[1,2], Ho-Hyun Nahm[1,2], Hasung Sim[1,2], Jaehong Jeong[1,2], T. G. Perring[3], Hyungje Woo[3,4], Kenji Nakajima[5], Seiko Ohira-Kawamura[5], Zahra Yamani[6], Y. Yoshida[7], H. Eisaki[7], S.-W. Cheong[8], A. L. Chernyshev[9], and Je-Geun Park[1,2*]

[1] Center for Correlated Electron Systems, Institute for Basic Science, Seoul 08826, Korea

[2] Department of Physics and Astronomy, Seoul National University, Seoul 08826, Korea

[3] ISIS Facility, STFC Rutherford Appleton Laboratory, Didcot OX11 0QX, United Kingdom

[4] Department of Physics, Brookhaven National Laboratory, Upton, New York 11973, USA

[5] Materials and Life Science Division, J-PARC Center, Japan Atomic Energy Agency, Tokai, Ibaraki 319-1195, Japan

[6] Chalk River Laboratories, National Research Council, Chalk River, Ontario K0J 1J0, Canada

[7] National Institute of Advanced Industrial Science and Technology, Tsukuba, Ibaraki 305-8565, Japan

[8] Department of Physics and Astronomy and Rutgers Center for Emergent Materials, Rutgers University, Piscataway, New Jersey 08854, USA

[9] Department of Physics and Astronomy, University of California, Irvine, California 92697, USA

* Corresponding author: jgpark10@snu.ac.kr

**When magnons and phonons, the fundamental quasiparticles of the solid, are coupled to one another, they form a new hybrid quasi-particle, leading to novel phenomena and interesting applications. Despite its wide-ranging importance, however, detailed experimental studies on the underlying Hamiltonian is rare for actual materials. Moreover, the anharmonicity of such magnetoelastic excitations remains largely unexplored although it is essential for a proper understanding of their diverse thermodynamic behaviour as well as intrinsic zero-temperature decay. Here we show that in noncollinear antiferromagnets, a strong magnon-phonon coupling can significantly enhance the anharmonicity, resulting in the creation of magnetoelastic excitations and their spontaneous decay. By measuring the spin waves over the full Brillouin zone and carrying out anharmonic spin wave calculations using a Hamiltonian with an explicit magnon-phonon coupling, we have identified a hybrid magnetoelastic mode in $(Y,Lu)MnO_3$ and quantified its decay rate and the exchange-striction coupling term required to produce it. Our work has wide implications for understanding of the spin-phonon coupling and the resulting excitations of the broad classes of noncollinear magnets.**

Spin and lattice constitute two of the four fundamental degrees of freedom in the solid: the other two are charge and orbital. In linearized models that account for many current understandings of the solid, excitations of spin and lattice: magnon of spin waves and phonon of lattice vibrations, are two principal examples of such quasiparticles [1]. Although there have been some experimental observations of cross-coupling between magnon and phonon, it is rather rare to actually observe and, more importantly, quantify the magnon-phonon coupling for real materials [2,3]. Nevertheless, it is essential for a proper understanding of their diverse thermodynamic behaviour as well as intrinsic zero-temperature decay [4]. Furthermore, it is generally believed that the magnon-phonon coupling is important for materials like multiferroic compounds, geometrically frustrated systems, spin-Peierls systems and Invar materials, to name only a few [5,6,7,8].

Quasiparticles like magnon and phonon, the cornerstone of modern condensed matter physics, are fundamentally the byproducts of linearized theories that ignore all the higher order terms than quadratic terms and neglect any conceivable interaction among the quasiparticles themselves. As such, they are considered to be stable, except for very few exceptions. For example, for classical spin systems without strong quantum fluctuations, magnon breakdown is thought to be unlikely for most of purposes. Therefore, observing and understanding how the otherwise stable quasiparticles break down in these unusual cases are the central theme of condensed matter physics.

One route leading to the breakdown of magnon and phonon is the cubic anharmonicities. Despite the general belief that this nonlinear magnon(-magnon) interaction is rather weak in real materials, recent insights gained mainly through theoretical studies suggest that things should be drastically different for certain cases, namely for noncollinear antiferromagnetic structures [4]. Unlike collinear magnetic structures that forbid the cubic anharmonicities, it was shown that such interaction is in principle allowed for noncollinear magnetic structures, such as the canonical 120° spin pattern in a 2D triangular lattice. There have since been several experimental reports [9,10] supporting these theoretical postulates. Nevertheless, the full details of the nonlinear interaction still needs to be worked out especially from experiments. We should also point out that this noncollinear magnetic order, in principle, allows a hitherto forbidden magnon-phonon coupling that has been less recognized among the community: the first order variation of the exchange energy with respect to transverse spin fluctuations is nonzero for a noncollinear magnetic order [11]. Because the O(3) symmetry is completely broken in the non-collinear ordered ground state of spins, coupling to phonons necessarily generates a coupling in which a magnon can convert *directly* into a phonon and vice versa. This is in contrast to the spin-lattice coupling in more conventional, collinear magnets, where the coupling usually respects the parity, which necessarily conserves the number of magnons, or allows creation (annihilation) of them only in pairs [12].

Here, we report the direct observation of the magnon-phonon coupling and the spontaneous decay of magneto-elastic excitations in the triangular antiferromagnets (Y,Lu)MnO$_3$. The full magnetic excitation spectra measured by inelastic neutron scattering experiments show clear deviations from the linear spin wave theory without the magnon-phonon coupling: an additional mode at high energies and the downward shift of the bottom mode at the Brillouin zone boundary. This is the most direct and stark evidence of the *linear* coupling of magnons and phonons, which, in turn, leads to enhancement of the anharmonic effects. We demonstrate that these experimental anomalies can only be fully resolved by incorporating the magnon-phonon coupling and carrying out the nonlinear spin wave analysis. We further reveal that the magneto-elastic excitation leads to significant broadening of the magnon spectra at the zone boundaries, originating from the decay of the magneto-elastic excitations into the two-magnon continuum.

Hexagonal rare-earth manganite RMnO$_3$ represents a good model system for geometrical frustration on a 2D triangular lattice: the nearest neighbour antiferromagnetic interaction between S=2 (Mn$^{3+}$) spins dominate, whilst the inter-layer interaction is relatively weak [13,14]. We should note that it also exhibits a very large spin-lattice coupling when it becomes magnetically ordered [5]. In a more recent work, we reported experimental evidence for a spontaneous magnon decay and a remarkably large spatial anisotropy for $Mn^{3+}$ ions in the exchange interactions for $LuMnO_3$. This was attributed to a structural distortion in which groups of three Mn atoms become more closely bound [9], such that the intratrimer $J_1$ exchange constants may differ from the intertrimer $J_2$ (see Fig. 1). Similar interpretations were shared by other groups too [15,16]. However, we should note that the large $J_1/J_2$ ratio of 6.4 obtained from fitting the data is inconsistent with the value of 1.15, obtained from first-principles calculations [17] that used the experimental atomic positions as reported by neutron diffraction [5]. This realization prompted us to handle both magnon and phonon as well as their cross coupling on an equal footing and to go beyond the linear spin wave analysis.

Fig.1 shows the spin waves measured at the MAPS beamline of the ISIS facility together with the theoretical dispersion relation calculated from the spin Hamiltonian by a linear spin wave theory using the following parameters (see the Supplementary Information): for $YMnO_3$, $J_1$=4, $J_2$=1.8, $D_1$=0.28, $D_2$=-0.02 meV: for $Y_{0.5}Lu_{0.5}MnO_3$, $J_1$=12.5, $J_2$=0.97, $D_1$=0.18, $D_2$=-0.018 meV: for $LuMnO_3$, $J_1$=9, $J_2$=1.4, $D_1$=0.28, $D_2$=-0.02 meV. Despite the apparent success of the linear spin wave calculations, there lies a critical failure: first, the downward curvature along the AB direction and second, the additional peaks at about 19 meV indicated by a red box in Fig.1c. But most importantly, here we have to use an unphysically large $J_1/J_2$ ratio in order to explain the additional high-energy peaks. Apart from the large $J_1/J_2$ ratio, this analysis of the linear spin waves has another drawback: which is that the calculated dynamical structure factor using the linear spin wave theory as shown in Figs. 1g & 1h always produces stronger intensity at the top mode of the spin waves than at the middle one, in marked contrast with the experimental data.

This discrepancy requires us to adopt a radically different approach to explain the measured full spin waves and go beyond the standard linear spin wave theory. One clue for how to address this problem can be taken from the physical properties: for example, our earlier neutron diffraction data revealed a giant spin-lattice coupling for $(Y,Lu)MnO_3$ [5]. This observation was subsequently corroborated by independent measurements on rare-earth hexagonal $RMnO_3$ [18-25]. More importantly, ultrasound measurements on $YMnO_3$ found marked softening in $C_{11}$ and $C_{66}$, supporting our view that there is a strong in-plane deformation below $T_N$ [26]. This observation naturally indicates the importance of magnon and in-plane phonon coupling. This conclusion is also backed up by theoretical calculations [17,27].

Following this idea of a large spin-lattice coupling in RMnO$_3$, we took a first-principles approach to the magnon-phonon coupling. First, we construct the following full model Hamiltonian, which couples the in-plane manganese vibrations directly to the spin system:

$$H = H_{spin} + \hbar \sum_{i=1}^{90} \omega_i b_k^\dagger b_k + \tilde{\alpha} \sum_{ij} \left( \boldsymbol{e}_{O_{ij}i} \cdot \boldsymbol{u}_i + \boldsymbol{e}_{O_{ij}j} \cdot \boldsymbol{u}_j \right) \boldsymbol{S}_i \cdot \boldsymbol{S}_j \qquad (1),$$

where $\boldsymbol{e}_{O_{ij}i}$ denotes the unit vector connecting the i-th manganese atom and the neighbouring oxygen atoms between the i-th and j-th manganese atoms as shown in Fig. SI2, $\tilde{\alpha}$ is the exchange striction, $\tilde{\alpha} = \partial J / \partial r$, which is naturally made into a dimensionless exchange-striction constant $\alpha = \tilde{\alpha} \cdot 2d/J$ [28], and $d$ is the Mn-O bond length at the equilibrium. Therefore, our Hamiltonian takes into account the modulations of the Mn-O bond length as a function of Mn displacements.

Before going into detailed discussion, we would like to make a general remark on the related issue. In cases when the spin rotational symmetry is broken completely in the ground state, i.e., when the spin structure is non-collinear, the Heisenberg term of the Hamiltonian provides a coupling of the transverse and longitudinal modes on neighboring sites. That is, in terms of the local site-dependent preferred spin direction of the ordered state, the coupling terms take the form of the type $S_i^z S_j^x$, etc. In the magnon language, they are quantized into the "odd" terms, producing linear $(a + a^\dagger)$ and cubic $(a^\dagger a a$, etc.) contributions. In equilibrium, the linear magnon term must vanish, leaving the anharmonic cubic magnon coupling to be the sole outcome, which is important for magnon decays. However, in the presence of coupling to phonons, the linear $(a + a^\dagger)$ terms is "activated", as the local atomic displacements $(u_i)$ violate the equilibrium conditions locally, hence the "direct" coupling of magnons and phonons.

To calculate the full dispersions of all 90 phonon modes for the unit cell with six formula units, we used a first-principles density functional theory (DFT). We show the full phonon dispersion curves for the three compounds as dashed lines in Figs. 2. We note that the calculated phonon DOS is in good agreement with the phonon spectra we measured using powder YMnO$_3$ and LuMnO$_3$ at the AMATERAS beamline of J-PARC (see the Supplementary Information). We then calculated the dynamical spin structure factor within the linear approximation by using the full Hamiltonian above with the explicit magnon-phonon coupling: we used the dimensionless exchange-striction coefficients of α~16-20.

The exchange-striction constant can also be estimated by using the pressure-dependence of the crystal structure and the antiferromagnetic transition reported for YMnO$_3$ [29,30]. Using the experimental data reported in Refs. 9 and 10, we came to an estimate of the dimensionless exchange striction α of 14 with the following formula: $\alpha = \frac{d(P_0)\partial T_N(P)/\partial P}{T_N(P_0)(\partial d(P)/\partial P)}$. Here, *d* is average Mn-O bond length, which is approximately one third of lattice constant *a*. The experimental parameters used in our estimate are $\frac{\partial T_N(P)}{\partial P} = 3$ K/GPa [29] and $\frac{\partial d(P)}{\partial P} = 0.0057$ Å/GPa [30]. Below we point out that our own data for the magnon excitation spectrum imply the value of the magneto-striction of the same order. This should be contrasted with the cuprates family, where estimates for an equivalent quantity are substantially smaller, α~2-7 [31,32].

By comparing with the experimental data, we obtained the best fitting results with the following sets of the parameters: $J_1=J_2=2.5$ meV, $D_1=0.28$ meV, $D_2=-0.02$ meV and α=16 for YMnO$_3$; $J_1=J_2=2.7$ meV, $D_1=0.28$ meV, $D_2=-0.02$ meV and α=20 for Y$_{0.5}$Lu$_{0.5}$MnO$_3$; $J_1=J_2=3$ meV, $D_1=0.28$ meV, $D_2=-0.02$ meV and α=16 for LuMnO$_3$. We note that according to the DFT calculations [17] the relative difference between $J_1$ and $J_2$ is theoretically about 10 – 20% at maximum. Therefore, we judge that this choice of $J_1=J_2$ in our analysis is good enough to capture the essential underlying physics of (Y,Lu)MnO$_3$, which is the magnon-phonon coupling.

The results shown as colour contour plot in Fig. 2 reproduce the overall features of the observed spectra of the experimental data in Fig. 1. It clearly shows that the high-energy signals located at 18 ~ 20 meV come from a direct coupling between the magnon and the optical phonons, i.e. a magneto-elastic mode. The intensity of this high energy magneto-phonon modes becomes stronger for Lu enriched compounds due to the larger Mn phonon DOS at this energy for the Lu enriched compounds, consistent with the experimental results shown in Fig. 1. This conclusion on the relevance of the magnon-phonon coupling for RMnO$_3$ is also supported by the fact that our calculated spin waves successfully explains the downward curvature of the bottom magnon branch along the AB direction. In fact, an estimate of the exchange striction from the splitting of the high-energy hybrid modes in Fig. 2 yields the values in the same range, α~10-20 (see Supplementary Information). We also note that our polarized neutron scattering data are in good agreement with our calculations (see Fig. SI4 of Supplementary Information).

Spontaneous decay of the hybrid mode: In addition to the magnon-phonon hybridization, noncollinear spin structures allow three magnon interactions as discussed above, which can lead to spontaneous magnon decay into two magnon states when the kinematic conditions are satisfied. The magneto-elastic excitations have, by definition, both magnon and phonon characters. Therefore, the above mechanism can also lead to the decay of magneto-elastic excitations inside the two quasi-particles continuum of magnon. Indeed, we observe significant broadening of the top mode in LuMnO$_3$ near the B and D points as shown in Fig. 3: less strong broadening has been seen for other two compounds.

In order to calculate the decay rate directly and compare it with the experimental data, we simplify the problem by assuming a dispersionless optical phonon mode at about 20 meV, where the strongest coupling has been observed in our data. Then our model Hamiltonian reads as follows:

$$H = J\sum_{ij}\left(S_i^x S_j^x + S_i^y S_j^y + \gamma S_i^z S_j^z\right) + \hbar\omega_0 \sum_k b_k^\dagger b_k + \tilde{\alpha}\sum_{ij}\left(\bm{e}_{O_{ij}i}\cdot \bm{u}_i + \bm{e}_{O_{ij}j}\cdot \bm{u}_j\right)\bm{S}_i\cdot \bm{S}_j$$

(2).

First, we calculate the dynamical spin structure factor by using a standard method. As shown in Fig. 3(a), despite the simplification, the calculation results reproduce well the experimental intensity along the C-B-D direction. For the calculations, we used the following set of the parameters: for YMnO$_3$ J=2.7 meV, γ=0.93, ℏω$_0$=17.5 meV, α=7.2; for LuMnO$_3$ J=3.2 meV, γ=0.93, ℏω$_0$=19.5 meV, α=8. Then the decay rate of the high-energy mode was calculated using an anharmonic spin wave theory within the 1/S approximation. The calculated results summarized in Fig.3 also show the significant linewidth broadening for the top mode near the B and D points only for LuMnO$_3$, consistent with the experimental results.

The reason for this is that in LuMnO$_3$, the combination of the higher energies of the magnon and magnetoelastic modes means that more decay channels, including two quasiparticle emission, are kinematically allowed. The different decay channels have different boundaries in the reciprocal space which also corresponds to logarithmic singularities in the decay rate [33], and the largest broadening is observed at momentum transfers where the single-quasiparticle dispersion crosses these boundaries, such as at the B point. In the case of YMnO$_3$, no such crossing occurs, so there are fewer decay channels available explaining why the observed linewidths remain narrow. Here we should stress that the single-magnon branches do not cross the line of singularities, whereas the magnetoelastic mode does, so a pure magnon decay is forbidden in this case. Similarly, the intrinsic decay rate of phonons is usually small due to a weak cubic anharmonicity. Thus the strong hybridization of magnons and phonons provides a new mechanism to enhancing the magnon decays.

Our studies using an exchange-striction model indicates that the deviations from the linear spin wave theory must be a common feature for other triangular antiferromagnets with a noncollinear magnetic order. Looking beyond the 2D triangular antiferromagnets, we believe that the idea of magnon-phonon coupling can also be important in the studies of a wide variety of the 3d transition-metal magnetic compounds. For example, similar analysis might shed a light on the investigation of spin phonon coupling mechanism and anharmonic effects in many other important noncollinear magnets that exhibit a rather large spin-lattice coupling, such as spinel [34] and invar materials [35], which should be dominated by the exchange-striction as discussed in Ref. [36]. This is in contrast to the previous studies on magnon phonon coupling focusing on materials with strong spin-orbit coupling such as rare-earth elements [37].

To summarize, we mapped out the spin waves and phonon excitations of (Y,Lu)MnO$_3$ over the Brillouin zone. By carrying out the spin wave calculations using the full Hamiltonian with both magnons and phonons on an equal footing as well as their coupling, we have not only demonstrated in our inelastic neutron scattering data a clear sign of magnon-phonon coupling, but also have quantified the coupling strength directly. Our work provides the rare experimental test and quantification of magnon-phonon coupling in real materials and opens a new window of opportunities in other materials such as 2D triangular lattice and other frustrated systems, where such couplings, hitherto hidden, have been long suspected.

**Acknowledgements**

We thank Kisoo Park and W. J. L. Buyers for useful discussions. The work at the IBS CCES was supported by the research programme of Institute for Basic Science (IBS-R009-G1). Experiment at J-PARC was carried out through the proposal number (2011B0064). The work of ALC was supported by the U.S. DOE, Office of Science, Basic Energy Sciences under Award # DE-FG02-04ER46174. The work at Rutgers University was supported by the DOE under Grant No. DOE: DE-FG02-07ER46382.

**Authors Contributions**

JGP conceived and supervised the project. HS, YY, HE and SWC grew the single crystals and HS

carried out bulk characterization. JO, MDL, JJ, TGP, HW, KN, SOK, ZY, and JGP were involved in the neutron scattering experiments. HHN carried out first-principles phonon calculation and JO made all the spin-wave calculations under the guidance of JGP and ALC. JGP, JO and ALC wrote the paper after discussion with all the authors.

**Competing financial interests**: The authors declare no competing financial interests.

**Methods**

Sample preparation: We synthesized powder samples using a solid state reaction method following the recipe as described in the literature [38]. We then grew single crystals of $Y_{1-x}Lu_xMnO_3$ (with typical size of 5×5×40 mm$^3$) by using a commercial optical floating zone furnace (Crystal Systems, Japan). Our subsequent powder and single crystal XRD confirmed that all our samples are prepared in high quality. We also measured the bulk properties (susceptibility and heat capacity) of all the samples to further confirm the quality by using a commercial set-up (MPMS5XL and PPMS9, Quantum Design USA).

Inelastic neutron scattering: Inelastic neutron scattering experiments have been performed on single crystal samples using the MAPS time-of-flight (ToF) spectrometer at ISIS, UK, and a triple axes spectrometer (TAS) with polarization analysis at Chalk River, Canada. In the ToF experiments, incident energies were chosen at 40 meV for $LuMnO_3$, 35 meV for $Y_{0.5}Lu_{0.5}MnO_3$, and 30 meV for $YMnO_3$ to adjust to the slight variations in the energy scales for each samples. The chopper frequency has been set to 250 Hz, which gave us a FWHM (full width at half maximum) energy resolution of 0.43 ~ 1.36 meV depending on the energy transfer. The measurements have been performed at 4 K for $YMnO_3$ and $Y_{0.5}Lu_{0.5}MnO_3$ and 13 K for $LuMnO_3$. We used the Horace program for our data analysis [39]. In the TAS experiments, spin polarized neutrons have been produced by using vertical focusing Heusler monochromator and analyser with the final energy fixed at 13.7 meV. In order to measure phonon DOS, inelastic neutron scattering experiments have also been performed on the powder samples with the incident energy of 42 meV using the AMATERAS beamline at J-PARC, Japan.

Theoretical calculations: We carried out first-principles calculations of phonon using a DFT+U method with U=4 eV. We used the PHONOPY code based on the force constant method [40]. And the force constants were constructed by means of a supercell approach based on the density functional perturbation theory (DFPT) [41], implemented in the VASP code [42]. Detailed discussion is given in the Supporting Information.

For the spin waves calculations, we used a rotating framework with the direction of easy axis anisotropy being rotated from parallel to perpendicular to the crystallographic axes. In order to make our calculations simpler and transparent, we ignored the interlayer exchange coupling as it is known to be more than 100 times smaller than the in-plane coupling [9,14]. Using this approximation, we can have the following minimal Hamiltonian: $J_1 \sum_{intra} \mathbf{S}_i \cdot \mathbf{S}_j + J_2 \sum_{inter} \mathbf{S}_i \cdot \mathbf{S}_j + D_1 \sum_i (S_i^z)^2 + D_2 \sum_i (S_i^n)^2$ (3), where $J_1$ and $J_2$ represent intra and inter trimer exchange constants and $D_1$ and $D_2$ are two magnetic anisotropies. We then calculated the spin wave dispersion using the standard linear spin wave theory [43]. We give detailed description of our spin wave calculations for the full Hamiltonian with the magnon and phonon coupling and the magnon-magnon nonlinear interaction in the Supporting Information.

Fig. 1 **Magnetic excitation spectra in (Y,Lu)MnO$_3$.** (a) A Mn-O layer in RMnO$_3$ forming a distorted two dimensional triangular antiferromagnet. (b) Inelastic neutron scattering data on LuMnO$_3$ summed over an energy window of [19.5, 20.5] meV. The arrows in (b) indicate the reciprocal points where the data shown in (c,d,e) are cut. (c,d,e) the inelastic neutron scattering data along the high symmetric directions (red circle and contour map) and fitted dispersion (black solid curve) for (c) YMnO$_3$ and (d) Y$_{0.5}$Lu$_{0.5}$MnO$_3$ (e) LuMnO$_3$ calculated by linear spin wave theory. (f,g,h) Calculated dynamical spin structure factors using the minimal spin Hamiltonian, equation (4) in the Supporting Information for (f) YMnO$_3$, (g) Y$_{0.5}$Lu$_{0.5}$MnO$_3$ and (h) LuMnO$_3$. For our simulations, we used the momentum and energy resolution of the instrument at the elastic line.

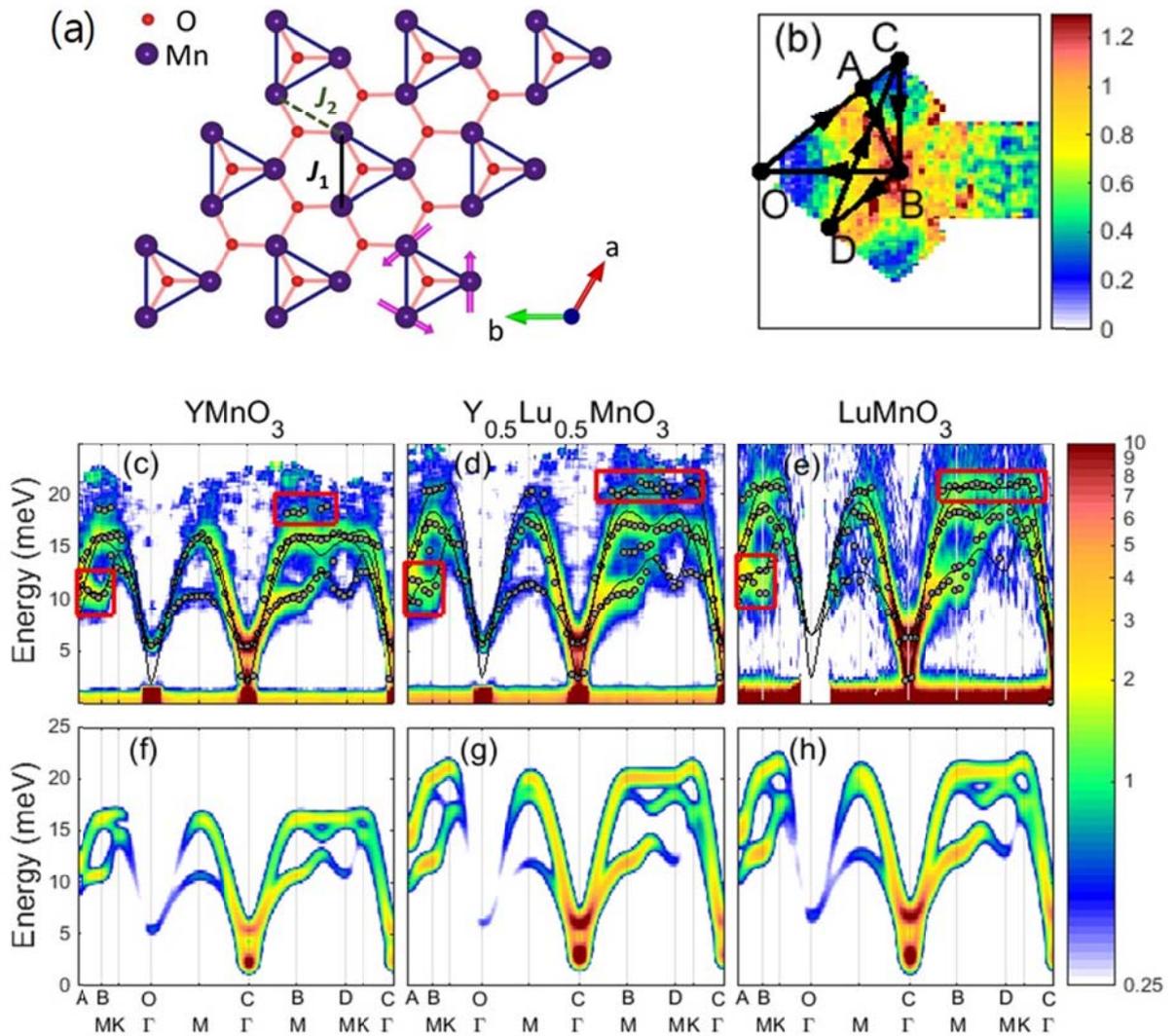

Fig. 2 **Calculated dynamical structure factor of magneto-elastic excitation**. The dynamical spin structure factor calculated from the full Hamiltonian, equation (1) (contour map) by taking into account the magnon-phonon coupling: the phonon dispersion curves (dotted lines) and the magnon dispersion without the coupling (solid lines) for (a) YMnO$_3$, (b) Y$_{0.5}$Lu$_{0.5}$MnO$_3$ and (c) LuMnO$_3$.

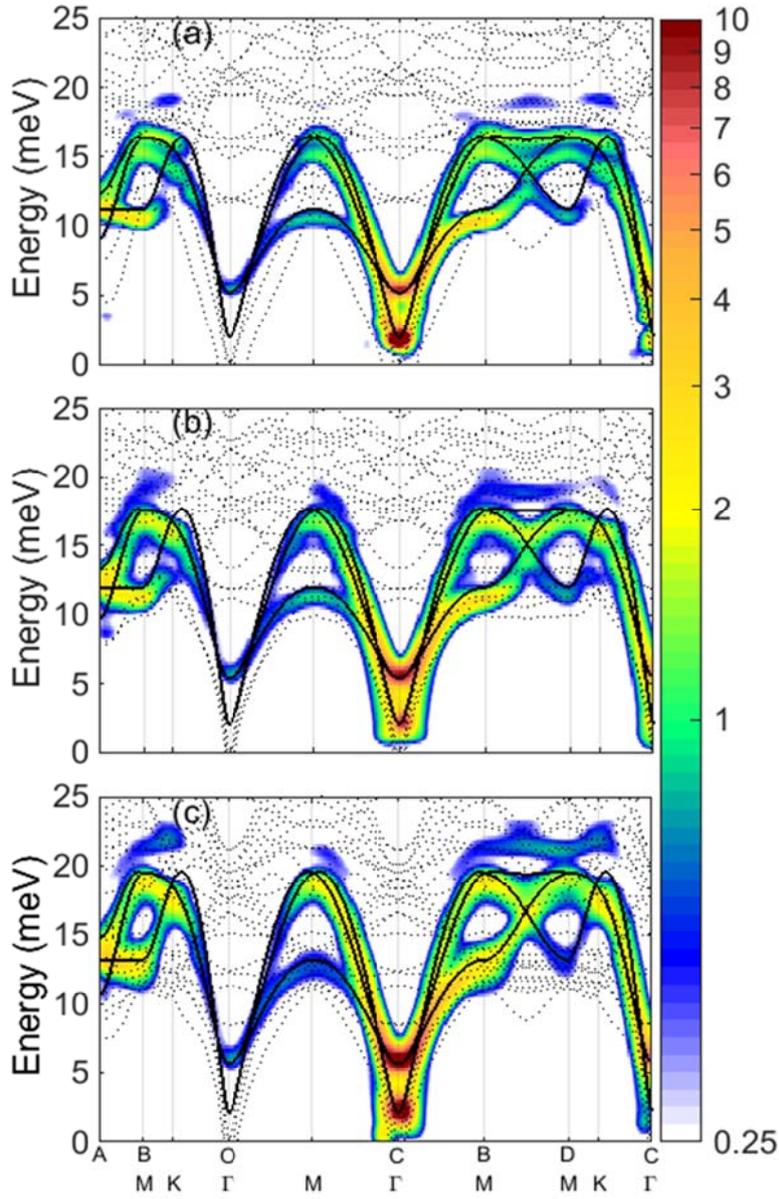

Fig. 3 **Linewidth broadening of magneto-elastic excitation.** (a) The neutron scattering data along the CBD direction (middle), the calculated dynamical structure factor within the linear spin wave theory (bottom), Observed linewidth broadening of the top mode (square) together with the calculated result from the 1/S approximation (line) and the experimental resolution (dashed line) (top) for YMnO$_3$ (left) and LuMnO$_3$ (right). (b) Observed linewidth of the top mode for LuMnO$_3$ (contour map) and (c) calculated intrinsic broadening of the top mode using the model Hamiltonian, equation (2). The experimental linewidth of the top mode was estimated by using multi Gaussian functions, and the typical results are shown in Fig. SI3 in the supporting material.

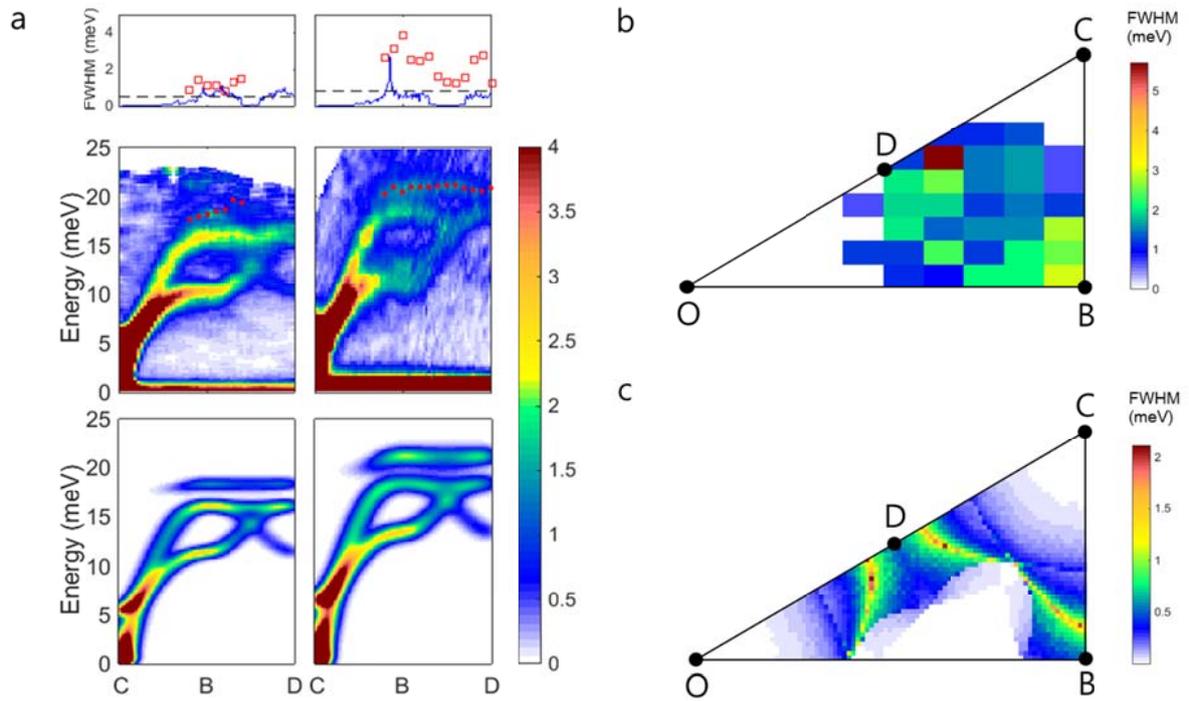